\documentclass[reqno,12pt]{article}
\usepackage{amsmath,amsfonts,amssymb,amsthm,amstext,amscd,eucal,xcolor}
\usepackage{color}
\usepackage[all]{xy}
\usepackage{mathtools}
\newcommand{\mathsym}[1]{{}} 
\usepackage[colorlinks=true,
            linkcolor=blue,
            urlcolor=blue,
            citecolor=blue]{hyperref}
\usepackage{enumerate}
\usepackage{enumitem}
\usepackage{cancel}
\usepackage{slashed}

\usepackage{graphicx}
\usepackage{bm}
\usepackage{cite,hyperref}
\usepackage{mathtools}
\usepackage{graphicx}
\usepackage[active]{srcltx}
\DeclareMathAlphabet{\pazocal}{OMS}{zplm}{m}{n}
\makeatletter \@addtoreset{equation}{section}

\makeatletter\renewcommand\section{\@startsection {section}{1}{\z@}%
                                   {-3.5ex \@plus -1ex \@minus -.2ex}
                                   {2.3ex \@plus.2ex}%
                                   {\normalfont\large\bfseries}}
\renewcommand\subsection{\@startsection{subsection}{2}{\z@}%
                                     {-3.25ex\@plus -1ex \@minus -.2ex}%
                                     {1.5ex \@plus .2ex}%
                                     {\normalfont\bfseries}}

\DeclareMathAlphabet{\mathcal}{OMS}{cmsy}{b}{n}

\makeatother

\newcommand{\email}[1]{\footnote{E-mail: \href{mailto:#1}{#1}}}

\begin{document}

\title{\bf\Large{Path integral analysis of the axial anomaly in Very Special Relativity } }

\author{ \bf{R.~Bufalo}\email{rodrigo.bufalo@ufla.br} $^{1}$ and \bf{M.~Ghasemkhani}\email{m\_ghasemkhani@sbu.ac.ir} $^{2}$  \vspace{0.3cm}\\
\textit{$^{1}$\small Departamento de F\'isica, Universidade Federal de Lavras,}\\
\textit{\small Caixa Postal 3037, 37200-900 Lavras, MG, Brazil}\\
\textit{$^{2}$\small Department of Physics, Shahid Beheshti University, 1983969411, Tehran, Iran }
}

\maketitle
\date{}

\begin{abstract}
In this note, we revise the problem of the axial anomaly in the framework of very special relativity following Fujikawa's path integral approach.
We show nonperturbatively that no VSR contribution is present in the path integral measure in $(3+1)$-dimensional spacetime.
Furthermore, we extend our results to $(1+1)$ dimensions, as well as to the two-dimesional curved spacetime.
\end{abstract}

\section{Introduction}

The study of anomalies in gauge theories is extremely important in demonstrating its consistency and physical properties, such as unitarity and renormalizability, as they are expected to cancel.
In order to verify whether a symmetry is preserved at the quantum level, one must analyze the respective Ward identity (or Slavnov-Taylor identity in the non-Abelian case).

Remarkably, in some special cases, the full structure of the anomaly can be calculated exactly at the one-loop order, and in the case of the chiral anomaly, it is given by the famous Adler-Bell-
Jackiw (ABJ) anomaly, justifying the $\pi_0 \to \gamma \gamma$ decay \cite{Bell:1969ts,Adler:1969gk,Jackiw:1986dr,bertlmann,fujikawa,Chaichian:2001da}.
As an another physical application of the anomalies, we can refer to the detected signatures of the chiral anomaly in the magneto-transport measurements in Weyl semimetals \cite{Zhang:2016ufu}.

Naturally, the models describing the physics beyond the standard model should fulfill such strong consistency requirements as well.
In this paper, we shall consider the problem of the anomalies in a particular class of the models presenting
Lorentz symmetry violation \cite{ref53,Bluhm:2005uj,Ellis:2005wr,AmelinoCamelia:2008qg}.
Contrary to the common lore, Lorentz violating effects are not necessarily related to Planck scale physics, it is also possible to formulate such class of models from a phenomenological group theory point of view.

A Lorentz violating framework, preserving all the basic elements of the special relativity, is the Cohen and Glashow very special relativity (VSR) \cite{Cohen:2006ky,Cohen:2006ir}.
The main aspect of the VSR proposal is that the laws of physics
are invariant under the (kinematical) subgroups of the Poincar\'e group.
In $(3+1)$--dimensional spacetime, there are two VSR subgroups, SIM(2) and HOM(2), that preserve the direction of a lightlike four-vector $n_{\mu}$ by scaling, transforming as $n \to e^{\varphi} n$ under boost in the z direction.
This feature implies that there is a preferred direction in the Minkowski spacetime, where the Lorentz violating terms can be constructed as ratios of contractions of the vector $n_{\mu}$ with other kinematical vectors, for instance $n^{\mu}/(n.p)$ \cite{Cohen:2006ky}.
The VSR approach has been extended to the gauge theories, where many interesting theoretical and phenomenological aspects of VSR effects have been extensively discussed \cite{Alfaro:2015fha,Upadhyay:2016hdj,Nayak:2016zed,Alfaro:2017umk,Bufalo:2017yxs,Alfaro:2019koq,Upadhyay:2019ghb,Bufalo:2019kea}.

Hence, since VSR gauge theories possess very interesting features related to the presence of gauge invariant massive modes, as well as nonlocal couplings \cite{Alfaro:2015fha}, it is a suitable Lorentz violating scenario to examine the chiral anomaly in the VSR electrodynamics.
This study could in principle signal some modification due to VSR effects of the (effective) coupling of the process $\pi_0 \to \gamma \gamma$ decay.
There were some works investigating the Adler-Bell-
Jackiw (ABJ) anomaly in the different Lorentz violating models \cite{Banerjee:2001un,Martin:2005gt,Sadooghi:2006sx,Arias:2007xt,Salvio:2008ta,Mariz:2015yua,AlMasri:2019uzr}, but not yet in the VSR context. 
Although, some aspects of the axial anomaly have been examined in the VSR-inspired Schwinger model \cite{Alfaro:2019snr}.

Recently, the reference \cite{Alfaro:2020zwo} presented a perturbative and a path integral analysis of the axial anomaly in the very special relativity (VSR) quantum electrodynamics, showing that no VSR contribution is present in the anomaly.
However, the author assumed some superfluous conditions in the analysis, such as ``light-cone gauge'', ``fixed A'' and ``large q'', which are actually unnecessary within Fujikawa's path integral approach \cite{bertlmann,fujikawa,Chaichian:2001da}.
Hence, in order to revise and generalize the path integral analysis of \cite{Alfaro:2020zwo}, our development follows strictly Fujikawa's sum regularization prescription to obtain the chiral Jacobian in a very elegant fashion \cite{bertlmann,fujikawa,Chaichian:2001da}.
Furthermore, contrary to \cite{Alfaro:2020zwo}, our analysis is as general as it can be, since it is nonperturbative, in the sense that no expansion is made in the nonlocal VSR terms.

We start Sec.~\ref{sec2} by reviewing the main features of the Fujikawa's approach, as well as presenting the main definitions in regard of very special relativity.
We focus great attention in presenting the regularization sum procedure since there lies the most important analysis involving the computation of the VSR effects into the Jacobian associated with the chiral symmetry, where we can clearly find how the VSR terms contribute.
Furthermore, we extend our results to the two-dimensional case.
In Sec.~\ref{sec3}, we generalize the analysis of the VSR contributions to the axial anomaly in curved spacetime, where we have considered by simplicity a two-dimensional curved spacetime.
In Sec.~\ref{conc}, we summarize the results, and present our final remarks.

\section{Fujikawa's approach to the axial anomaly}
\label{sec2}

In order to investigate the axial anomaly in the context of path integral approach, it is necessary to establish the path integral measure accurately \cite{bertlmann,fujikawa,Chaichian:2001da}. We start with the Euclidean VSR path integral
\begin{equation}
Z\left[A\right]=\int D\overline{\psi}D\psi\exp\left[\int d^{4}x~\bar{\psi}\left(i\slashed{\nabla}-m_{e}\right)\psi\right],
\end{equation}
where $\nabla_{\mu}=D_{\mu}-\frac{m^{2}n_{\mu}}{2\left(n.D\right)}$ is defined as the VSR covariant derivative with $D_{\mu}=\partial_{\mu}-ieA_{\mu}$, and $m$ is the VSR parameter.
\footnote{ Although a gauge fixing is allowed in this gauge-invariant approach, we shall show that contrary to \cite{Alfaro:2020zwo}, where the light-cone gauge $n.A= 0$ was used in order to simplify the operator $\nabla$ to $\nabla_{\mu}=D_{\mu}-\frac{m^{2}n_{\mu}}{2\left(n.\partial\right)}$ (removing all nonlocal couplings), we can perform the whole analysis in path integral without any gauge fixing.} 
We would like to analyze the path integral measure behavior under local chiral transformations
\begin{equation}
\psi\to e^{i\beta\left(x\right)\gamma_{5}}\psi,\quad\overline{\psi}\to\overline{\psi}e^{i\beta\left(x\right)\gamma_{5}}, \label{eq2}
\end{equation}
and then determine whether VSR effects can modify the well-known axial anomaly \cite{bertlmann,fujikawa,Chaichian:2001da}.

In this sense, we decompose the spinors into eigenfunctions of the
Dirac operator as follows
\begin{equation} \label{eq3}
\psi\left(x\right)=\sum_{n}a_{n}\varphi_{n}\left(x\right),\quad\bar{\psi}\left(x\right)
=\sum_{m}\varphi_{m}^{\dagger}\left(x\right)\overline{b}_{m},
\end{equation}
where the coefficients $a_{n}$ and $\overline{b}_{m}$ are independent
Grassmann variables. The hermitian Dirac operator $\nabla$ satisfies
\begin{equation}
\slashed{\nabla}\varphi_{n}\left(x\right)=\lambda_{n}\varphi_{n}\left(x\right),\label{eq13}
\end{equation}
where $\lambda_{n}$'s are real eigenvalues of $\nabla$, the set of
eigenfunctions is orthonormal and complete
\begin{align}
\int d^{4}x~\varphi_{m}^{\dagger}\left(x\right)\varphi_{n}\left(x\right) & =\delta_{mn},\nonumber \\
\sum_{n}\varphi_{n}\left(x\right)\varphi_{n}^{\dagger}\left(y\right) & = \delta^{(4)}\left(x-y\right),  \label{eq14}
\end{align}
The first point to emphasize is that this basis set allows us to diagonalize
\begin{equation}
\bar{\psi}\left(i\slashed{\nabla}-m_{e}\right)\psi=\sum_{n}\left(i\lambda_{n}-m_{e}\right)\overline{b}_{n}a_{n}.
\end{equation}
Hence, we realize that under the decomposition \eqref{eq3} the fermionic path integral measure can be expressed as  $
D\overline{\psi}D\psi  =\prod_{n}d\overline{b}_{n}da_{n}$.
In the next step, we shall consider how the fermionic fields behave under the local chiral transformations  \eqref{eq2}.
In particular, we are interested in analyzing how the fermionic path
integral measure behaves under these infinitesimal transformations.
Applying these considerations,we find the following transformation law for the coefficients of \eqref{eq3}
\begin{equation}
\left(\begin{array}{c}
a_{m}'\\
\overline{b}_{m}'
\end{array}\right)=\sum_{k}C_{mk}\left(\begin{array}{c}
a_{k}\\
\overline{b}_{k}
\end{array}\right),
\end{equation}
with the transformations matrix
\begin{equation}
C_{mk}=\delta_{mk}+i\int d^{4}x~\varphi_{m}^{\dagger}\left(x\right)\beta\left(x\right)\gamma_{5}\varphi_{k}\left(x\right).
\end{equation}
Finally, from these results, we can write the transformation for the
path integral measure as
\begin{align}
D\overline{\psi}'D\psi' & =\prod_{n}d\overline{b}_{n}'da_{n}' \equiv \pazocal{J}\left[\beta\right]D\overline{\psi}d\psi,
\end{align}
where we have identified the Jacobian  $\pazocal{J}\left[\beta\right]=\left(\det C\right)^{-2} $.
This functional can also be cast as
\begin{align} \label{eq5}
\pazocal{J}\left[\beta\right]  =\exp\Bigg(-2i\int d^{4}x~\beta\left(x\right)\sum_{m}\varphi_{m}^{\dagger}\left(x\right)\gamma_{5}\varphi_{m}\left(x\right)\Bigg),
\end{align}
which form is valid for infinitesimal $\beta\left(x\right)$.
However, this sum is not well defined, actually
\begin{equation}
\sum_{m}\varphi_{m}^{\dagger}\left(x\right)\gamma_{5}\varphi_{m}\left(x\right)= \text{tr}\left(\gamma_{5}\right)\delta\left(0\right),
\end{equation}
where we have used the completeness relation of the eigenfunctions  \eqref{eq14}.
Hence, it is necessary to regularize the sum in \eqref{eq5}, and we follow Fujikawa's prescription \cite{fujikawa}.
As we will show below, the correct use of the Fujikawa's regularization approach leads to an unambiguous analysis of the axial anomaly in the VSR setting.

In the regularization procedure, we shall consider an arbitrary function that is smooth and decreasing sufficiently rapidly at infinity
\begin{equation}
f\Big(\frac{\lambda_{m}^{2}}{\Lambda^{2}}\Big),
\end{equation}
with $\Lambda\to\infty$ and
\begin{align}
f\left(\infty\right) & =f'\left(\infty\right)=f''\left(\infty\right)=\cdots=0,\nonumber \\
f\left(0\right) & =1.
\end{align}
Hence, since the Dirac operator is now regularized, we can express the eigenfunctions in Fourier components, i.e. plane waves, and thus we can compute the Jacobian \eqref{eq5} explicitly.

First, we should consider
\begin{align}
\sum_{m}\varphi_{m}^{\dagger}\left(x\right)\gamma_{5}\varphi_{m}\left(x\right)  & =\lim_{\Lambda\to\infty}\int\frac{d^{4}q}{\left(2\pi\right)^{4}}\text{tr}\Big[e^{-iq.x}\gamma_{5}
f\Big(\frac{\slashed{\nabla}^{2}}{\Lambda^{2}}\Big)
e^{iq.x}\Big] \cr
& =\lim_{\Lambda\to\infty}\int\frac{d^{4}q}{\left(2\pi\right)^{4}}\text{tr}\Big[e^{-iq.x}\gamma_{5}
f \Big(\frac{\nabla_{\mu}\nabla^{\mu}}{\Lambda^{2}}+\frac{1}{4\Lambda^{2}}\left[\gamma^{\mu},\gamma^{\nu}\right]
\left[\nabla_{\mu},\nabla_{\nu}\right]\Big)e^{iq.x}\Big],
 \label{eq6}
\end{align}
where we have decomposed the Dirac operator as usual.
Furthermore, the effect of moving the plane waves is a shift in the differential operators
\begin{equation}
e^{-iq.x}f\left(\partial_{\mu}\right)e^{iq.x}=f\left(\partial_{\mu}+iq_{\mu}\right).
\end{equation}
Hence, we observe that under this change we have $D\to D+iq$. Thus, the VSR covariant derivative is written as
\begin{equation}
\nabla_{\mu}'=D_{\mu}+iq_{\mu}-\frac{m^{2}n_{\mu}}{2\left(n.D+in.q\right)}.
\end{equation}
This implies that the argument of the function $f(z)$ in \eqref{eq6} takes the form
\begin{align} \label{eq15}
z & =-q_{\mu}q^{\mu}+\frac{2iq_{\mu}\left[D^{\mu}-\frac{m^{2}n^{\mu}}{2\left(n.D+i(n.q)\Lambda\right)}\right]}{\Lambda}+\frac{\left(D_{\mu}-\frac{m^{2}n_{\mu}}{2\left(n.D+i(n.q)\Lambda\right)}\right)\left(D^{\mu}-\frac{m^{2}n^{\mu}}{2\left(n.D+i(n.q)\Lambda\right)}\right)}{\Lambda^{2}}\nonumber \\
 & +\frac{1}{4\Lambda^{2}}\left[\gamma^{\mu},\gamma^{\nu}\right]\left(\left[D_{\mu},D_{\nu}\right]-\frac{m^{2}n_{\nu}}{2}
 \left[D_{\mu},\frac{1}{\left(n.D+i(n.q)\Lambda\right)}\right]-\frac{m^{2}n_{\mu}}{2}
 \left[\frac{1}{\left(n.D+i(n.q)\Lambda\right)},D_{\nu}\right]\right).
\end{align}
Here, it is worth to formulate some comments as follows about this expression:

\begin{enumerate}

\item We have performed the rescaling in the momentum $q\to\Lambda q$, which allows gathering the relevant contributing terms of eq.\eqref{eq6};

\item  The $m^{4}$ term vanishes, since it is proportional to the product of a symmetric tensor $n_{\mu}n_{\nu}$ with $\left[\gamma^{\mu},\gamma^{\nu}\right]$;

\item Since $q$ is a fixed/constant 4-vector, its commutator with the derivative $D_\mu $ is zero, i.e. $\left[D_{\mu}, q_\nu \right]=0$;

\item In regard to the commutator $\left[D_{\mu},\frac{1}{\left(n.D+i(n.q)\Lambda\right)}\right]$, it vanishes identically because in the limit $\Lambda\to\infty$ the nonlocal factor goes as $\frac{1}{\left(n.D+i(n.q)\Lambda\right)}\sim\frac{1}{\left(i(n.q)\Lambda\right)}$,
i.e. $\left[D_{\mu},\frac{1}{\left(n.D+i(n.q)\Lambda\right)}\right]\sim\left[D_{\mu},\frac{1}{\left(i(n.q)\Lambda\right)}\right]=0$.

\end{enumerate}

Now, we expand the content of the function $f\left(z\right)$ around the value $y=-q_{\mu}q^{\mu}=q^{2}$.
Furthermore, only the terms proportional to $1/\Lambda^{4}$ or larger survive in the limit
$\Lambda\to\infty$. Also, looking at the trace of the $\gamma$-matrices
with $\gamma_{5}$, only terms with four or more $\gamma$-matrices
have nonvanishing traces. Hence, we find
\begin{align} \label{eq10}
\sum_{m}\varphi_{m}^{\dagger}\left(x\right)\gamma_{5}\varphi_{m}\left(x\right) & =\lim_{\Lambda\to\infty}\Lambda^{4}\int\frac{d^{4}q}{\left(2\pi\right)^{4}}\text{tr}\left[\gamma_{5}
\Big(f\left(y\right)+wf'\left(y\right)+\frac{w^{2}}{2!}f''\left(y\right)+\cdots\Big)\right],
\end{align}
where $w$ is defined as
\begin{align} \label{eq60}
w & =\frac{2iq_{\mu}\left[D^{\mu}-\frac{m^{2}n^{\mu}}{2\left(n.D+i(n.q)\Lambda\right)}\right]}
{\Lambda}+\frac{\left(D_{\mu}-\frac{m^{2}n_{\mu}}{2\left(n.D+i(n.q)\Lambda\right)}\right)
\left(D^{\mu}-\frac{m^{2}n^{\mu}}{2\left(n.D+i(n.q)\Lambda\right)}\right)}{\Lambda^{2}}
+\frac{1}{2\Lambda^{2}}\gamma^{\mu}\gamma^{\nu}\left(-ie\right)F_{\mu\nu},
\end{align}
where $F_{\mu\nu}=\partial_{\mu}A_{\nu}-\partial_{\nu}A_{\mu} = -\frac{1}{ie}\left[D_{\mu},D_{\nu}\right]$.
Although we have some terms of $w$ that survive in the limit $\Lambda\to\infty$ in the Taylor's expansion \eqref{eq10} but they all vanish due to trace identities: $\text{tr}(\gamma_5 )=\text{tr}(\gamma_5 \gamma^\mu\gamma^\nu ) =0$ and $\text{tr}(\gamma_5 \gamma^\mu\gamma^\nu\gamma^\alpha\gamma^\beta)  =-4\epsilon^{\mu\nu\alpha\beta}$.
 The only remaining term that survives after the regulator limit and taking the trace is 
\begin{align}
\sum_{m}\varphi_{m}^{\dagger}\left(x\right)\gamma_{5}\varphi_{m}\left(x\right)&=\lim_{\Lambda\to\infty}\Lambda^{4}\int\frac{d^{4}q}{\left(2\pi\right)^{4}}\text{tr}\left[\gamma_{5}\frac{1}{2!}\left(\frac{1}{2\Lambda^{2}}\gamma^{\mu}\gamma^{\nu}\left(-ie\right)F_{\mu\nu}\right)^{2}f''\left(y\right)\right]\cr
&=-\frac{e^{2}}{8}\text{tr}\left[\gamma_{5}\gamma^{\mu}\gamma^{\nu}\gamma^{\alpha}\gamma^{\beta}\right]F_{\mu\nu}F_{\alpha\beta}\int\frac{d^{4}q}{\left(2\pi\right)^{4}}f''\left(y\right), \label{eq20}
\end{align}
which corresponds exactly to the usual singlet axial anomaly expression
\begin{align}
\sum_{m}\varphi_{m}^{\dagger}\left(x\right)\gamma_{5}\varphi_{m}\left(x\right) & =\frac{e^{2}}{32\pi^{2}}\epsilon^{\mu\nu\alpha\beta}F_{\mu\nu}F_{\alpha\beta},
\end{align}
where we have made use of (obtained through integration by parts)
\begin{equation}
\int\frac{d^{4}q}{\left(2\pi\right)^{4}}f''\left(q^{2}\right)=-\frac{1}{16\pi^{2}}\int dq^{2} f'\left(q^{2}\right)=\frac{1}{16\pi^{2}}.
\end{equation}
Accordingly, the Jacobian of the path integral measure \eqref{eq5} in $(3+1)$ dimensions is given by
\begin{equation}
\pazocal{J}\left[\beta\right]=\exp\left(-\frac{ie^{2}}{16\pi^{2}}\int d^{4}x~\beta\left(x\right)\epsilon^{\mu\nu\alpha\beta}F_{\mu\nu}F_{\alpha\beta}\right).
\end{equation}
Hence, we have shown that under the correct use of the Fujikawa's regularization prescription, no VSR corrections are found for the axial anomaly in the path integral approach.

\subsection{Two-dimensional axial anomaly in VSR}

An import remark about the previous analysis in the evaluation of the operator $\left[\gamma^{\mu},\gamma^{\nu}\right]
\left[\nabla_{\mu},\nabla_{\nu}\right]$ in \eqref{eq15} is in order. The considerations and main results showing the absence of VSR effects are valid for any spacetime dimension.
As a matter of fact, the only actual change is the integral measure.
In the case of $(1+1)$ spacetime we have that eq.~\eqref{eq10} should be written as
\footnote{Actually, the Lorentz group in two dimensions i.e. SO(1,1) does not have any VSR subgroup, since it has only one generator, one boost. However, one can make use of a null-vector $n_\mu =(1,1)$ and construct a VSR-like QED action \cite{Alfaro:2019snr} with the same structure for the covariant derivative $\nabla_{\mu}=D_{\mu}-\frac{m^{2}n_{\mu}}{2\left(n.D\right)}$.}
\begin{align} \label{eq100}
\sum_{m}\varphi_{m}^{\dagger}\left(x\right)\gamma_{5}\varphi_{m}\left(x\right) & =\lim_{\Lambda\to\infty}\Lambda^{2}\int\frac{d^{2}q}{\left(2\pi\right)^{2}}\text{tr}\left[\gamma_{5}
\Big(f\left(y\right)+wf'\left(y\right)+\frac{w^{2}}{2!}f''\left(y\right)+\cdots\Big)\right],
\end{align}
with $w$ given by  \eqref{eq60}.
Another difference is the result $\text{tr}(\gamma_5 \gamma^\mu\gamma^\nu ) =-2\epsilon^{\mu\nu}$, that comes from the identity $\gamma_\mu\gamma_5=\epsilon_{\mu\nu}\gamma^\nu$, with $\epsilon_{01}=1$ (here, we follow the notation of \cite{bertlmann}).
Making use of these results we find that only the linear term in $w$ contributes in eq.~\eqref{eq100}, and the regularized sum is given by
\begin{align}
\sum_{m}\varphi_{m}^{\dagger}\left(x\right)\gamma_{5}\varphi_{m}\left(x\right)&=\lim_{\Lambda\to\infty}\Lambda^{2}\int\frac{d^{2}q}{\left(2\pi\right)^{2}}\text{tr}\left[\gamma_{5} \left(\frac{1}{2\Lambda^{2}}\gamma^{\mu}\gamma^{\nu}\left(-ie\right)F_{\mu\nu}\right) f'\left(y\right)\right]\cr
&=-\frac{ie}{2}\text{tr}\left[\gamma_{5}\gamma^{\mu}\gamma^{\nu}\right]F_{\mu\nu}\int\frac{d^{2}q}{\left(2\pi\right)^{2}}f'\left(q^{2}\right),
\end{align}
and using
\begin{equation}
\int\frac{d^{2}q}{\left(2\pi\right)^{2}}f'\left(q^{2}\right) = - \frac{1}{4\pi}
\end{equation}
we obtain
\begin{align}
\sum_{m}\varphi_{m}^{\dagger}\left(x\right)\gamma_{5}\varphi_{m}\left(x\right)  = -\frac{ie}{4\pi }\epsilon_{\mu\nu}F^{\mu\nu},
\end{align}
that corresponds to the usual axial anomaly in two-dimensions.

\section{Axial anomaly in curved spacetime}
\label{sec3}

As an additional analysis of the axial anomaly in VSR, we can extend our discussion to the case of chiral anomalies associated with the (background) gravitational field.
In regard to the formal development presented above to determine the functional Jacobian \eqref{eq5}, the main change in the case of curved spacetime is the spacetime measure $ d^{\omega}x \to \sqrt{-g}~ d^{\omega}x$ \cite{fujikawa}.

Now, about VSR in curved spacetime, there are some formal developments and applications found in the literature \cite{Alvarez:2008uy,Muck:2008bd}, where the Lorentz violating effects are studied in terms of (spinor) auxiliary fields $(\lambda, \chi)$
\begin{equation}
\pazocal{L} = i \bar{\psi} \gamma^\mu\pazocal{D}_{\mu}\psi +i \bar{\chi} (n.\pazocal{D})\lambda +i \bar{\lambda} (n. \pazocal{D})\chi +im \bar{\chi} (\gamma. n) \psi + im  \bar{\lambda}\psi +h.c.
\end{equation}
to avoid the presence of explicit non-local factors.
Furthermore, in this formulation $n^\mu$ is a recurrent null vector field satisfying $\pazocal{D}_{\mu}n^\nu =0$.
However, the construction of VSR invariant in curved spacetime is not a closed matter, we shall consider as a matter of discussion the results of \cite{Alvarez:2008uy} and assume that the functional form of the fermionic minimal coupling is $i\bar{\psi} \gamma^\mu \nabla_{\mu} \psi$, with the VSR covariant derivative defined as $\nabla_{\mu}=\pazocal{D}_{\mu}-\frac{m^{2}n_{\mu}}{2\left(n.\pazocal{D}\right)}$, but now with new couplings $\pazocal{D}_{\mu}=\partial_{\mu}-ieA_{\mu} +\frac{1}{2}\omega_{\mu}^{ab}\sigma_{ab}\equiv  D_\mu +\frac{1}{2}\Omega_{\mu}^{ab}\sigma_{ab}$, where $\omega_{\mu}^{ab}$ is the usual spin connection \cite{Alvarez:2008uy}.
Furthermore, $\sigma_{ab}$ stands for the representation appropriate for the field that the derivative acts on, e.g. in the case of the usual components $\omega_{\mu}^{ab}$ acting on a fermionic field, we have that $\sigma_{ab}=\frac{1}{4}[\gamma_a,\gamma_b]$.

By simplicity, we shall consider a two-dimensional curved spacetime to develop our analysis.
Once again, the matrix structure of the regularized sum is very similar to \eqref{eq100}. Thus, we shall focus only on the evaluation of the commutator  $\left[\nabla_{\mu},\nabla_{\nu}\right]$.
In this case, we have that
\begin{align} \label{eq150}
e^{-iq.x}\left[\nabla_{\mu},\nabla_{\nu}\right]e^{iq.x}&=  \left[\pazocal{D}_{\mu},\pazocal{D}_{\nu}\right] \cr
&-\frac{m^{2}n_{\nu}}{2}
 \left[\pazocal{D}_{\mu},\frac{1}{\left(n.\pazocal{D}+i(n.q)\Lambda\right)}\right]-\frac{m^{2}n_{\mu}}{2}
 \left[\frac{1}{\left(n.\pazocal{D}+i(n.q)\Lambda\right)},\pazocal{D}_{\nu}\right],
\end{align}
where the rescaling in the momentum $q\to\Lambda q$ was performed.
We can make use of the comments 3. and 4. presented after eq.~\eqref{eq15} to show that the last two terms of  \eqref{eq150} vanish identically.
Because in the limit $\Lambda\to\infty$ the nonlocal factor goes as $\lim\limits_{\Lambda\to\infty} \frac{1}{\left(n.\pazocal{D}+i(n.q)\Lambda\right)}\sim \lim\limits_{\Lambda\to\infty} \frac{1}{\left(i(n.q)\Lambda\right)}$,
i.e.
\begin{equation}
\lim\limits_{\Lambda\to\infty}\left[\pazocal{D}_{\mu},\frac{1}{\left(n.\pazocal{D}+i(n.q)\Lambda\right)}\right]\sim \lim\limits_{\Lambda\to\infty}\left[\pazocal{D}_{\mu},\frac{1}{\left(i(n.q)\Lambda\right)}\right]=0.
\end{equation}
We then find that
\begin{equation}
e^{-iq.x}\left[\nabla_{\mu},\nabla_{\nu}\right]e^{iq.x}=  \left[\pazocal{D}_{\mu},\pazocal{D}_{\nu}\right],
\end{equation}
showing that, once again, no VSR contribution is present in the axial anomaly in the two-dimensional QED, even in a curved spacetime.
As a matter of fact, the chiral anomaly of the two-dimensional QED in curved spacetime is the same as in the flat spacetime \cite{Barcelos-Neto:1985yfn}.

\section{Final remarks}
\label{conc}

In this paper, we have analyzed the four-dimensional chiral anomaly for Dirac fermions in the context of the very special relativity using the path integral approach, where we have found no presence of VSR effects in the Jacobian functional.
Although some points were dealt in ref.\cite{Alfaro:2020zwo}, we believe that the treatment in \cite{Alfaro:2020zwo} is somehow limited
because it makes use of unnecessary conditions such as ``light-cone gauge'' (gauge-fixing), ``fixed A'' and ``large q''.
Moreover, the main issue is the improper procedure in the regularization of the ill-defined sum in eq.\eqref{eq5}.

On the other hand, our analysis is as general as it can be since
it holds nonperturbatively.
As we have shown above, the correct application of Fujikawa's sum regularization procedure (in terms of the limit $\Lambda\to\infty$) is the key element to obtain the axial anomaly in the path integral formalism.
Mainly because it allows us to collect naturally the contributing terms in the Taylor's series  \eqref{eq10}.

 In addition, we have extended our discussion to the case of the two-dimensional chiral anomaly associated with the (background) gravitational field, showing again that no VSR effect is present.
In summary, as long as the VSR effects have the ``usual'' nonlocal form $\nabla_{\mu}=\pazocal{D}_{\mu}-\frac{m^{2}n_{\mu}}{2\left(n.\pazocal{D}\right)}$, independently of the functional form of the gauge connection (electromagnetic and gravitational part) of the covariant derivative $\pazocal{D}_{\mu}$, it can always be removed in the evaluation of the path integral Jacobian.

 Since there is a close relation among the ABJ anomaly with the $\pi_0 \to \gamma +\gamma$ decay \cite{Adler:1969gk,Jackiw:1986dr}, it is a natural extension to examine the VSR effects upon this type of anomalous processes and find how the VSR  nonlocal effects contribute to its decay width.
This is currently under development and will be reported elsewhere.

 \subsection*{Acknowledgements}

We are grateful to M. Chaichian and A. Tureanu for their encouragements to present this paper.
R.B. acknowledges partial support from Conselho
Nacional de Desenvolvimento Cient\'ifico e Tecnol\'ogico (CNPq Projects No. 305427/2019-9 and No. 421886/2018-8) and Funda\c{c}\~ao de
Amparo \`a Pesquisa do Estado de Minas Gerais (FAPEMIG Project No. APQ-01142-17).

\global\long\def\link#1#2{\href{http://eudml.org/#1}{#2}}
 \global\long\def\doi#1#2{\href{http://dx.doi.org/#1}{#2}}
 \global\long\def\arXiv#1#2{\href{http://arxiv.org/abs/#1}{arXiv:#1 [#2]}}
 \global\long\def\arXivOld#1{\href{http://arxiv.org/abs/#1}{arXiv:#1}}



\begin{thebibliography}{99}

\bibitem{Bell:1969ts}
J.~S.~Bell and R.~Jackiw,
``\textit{A PCAC puzzle: $\pi^0 \to \gamma \gamma$ in the $\sigma$ model},''
\doi{10.1007/BF02823296}{Nuovo Cim. A \textbf{60}, 47-61 (1969)}.

\bibitem{Adler:1969gk}
S.~L.~Adler,
``\textit{Axial vector vertex in spinor electrodynamics},''
\doi{10.1103/PhysRev.177.2426}{Phys. Rev. \textbf{177}, 2426-2438 (1969)}.

\bibitem{Jackiw:1986dr}
R.~Jackiw,
``Field theoretic investigations in current algebra,''
in: Current Algebra and Anomalies, S.B. Treiman, R. Jackiw, B. Zumino and E. Witten (eds.), p.81, and p. 211, World Scientific, Singapore.
\bibitem{bertlmann}
  R.A.~Bertlmann,
  ``Anomalies in Quantum Field Theory,''
International Series of Monographs on Physics 91, Oxford University Press (2001).

\bibitem{fujikawa}
  K.~Fujikawa and H. Suzuki,
  ``Path Integrals and Quantum Anomalies,'' International Series of Monographs on Physics 122, Oxford University Press (2004).

\bibitem{Chaichian:2001da}
M.~Chaichian and A.~Demichev,
``Path integrals in physics. Vol. 2: Quantum field theory, statistical physics and other modern applications,''
CRC Press (2001).

\bibitem{Zhang:2016ufu}
C.~Zhang, S.~Y.~Xu, I.~Belopolski, Z.~Yuan, Z.~Lin, B.~Tong, N.~Alidoust, C.~C.~Lee, S.~M.~Huang and T.~R.~Chang, \textit{et al.}
``\textit{Signatures of the Adler-Bell-Jackiw chiral anomaly in a Weyl Fermion semimetal},''
  \doi{10.1038/ncomms10735}{Nature Commun. \textbf{7}, 0735 (2016)},
\arXiv{1601.04208}{cond-mat.mtrl-sci}.

    \bibitem{ref53}
  D.~Mattingly,
  ``Modern tests of Lorentz invariance,''
     \doi{10.12942/lrr-2005-5}{Living Rev.\ Rel.\  {\bf 8} (2005) 5},
      \arXivOld{gr-qc/0502097}.



    \bibitem{Bluhm:2005uj}
  R.~Bluhm,
 ``Overview of the SME: Implications and phenomenology of Lorentz violation,''
  \doi{10.1007/3-540-34523-X_8}{Lect.\ Notes Phys.\  {\bf 702} (2006) 191},
    \arXivOld{hep-ph/0506054}.
\bibitem{Ellis:2005wr}
  J.~R.~Ellis, N.~E.~Mavromatos, D.~V.~Nanopoulos, A.~S.~Sakharov and E.~K.~G.~Sarkisyan,
  ``Robust limits on Lorentz violation from gamma-ray bursts,''
    \doi{10.1016/j.astropartphys.2006.04.001}{Astropart.\ Phys.\  {\bf 25} (2006) 402},
    \arXivOld{astro-ph/0510172}.
 \bibitem{AmelinoCamelia:2008qg}
  G.~Amelino-Camelia,
  ``Quantum-Spacetime Phenomenology,''
  \doi{10.12942/lrr-2013-5}{Living Rev.\ Rel.\  {\bf 16} (2013) 5},
  \arXiv{0806.0339}{gr-qc}.

\bibitem{Cohen:2006ky}
A.~G.~Cohen and S.~L.~Glashow,
  ``\textit{Very special relativity},''
  \doi{10.1103/PhysRevLett.97.021601}{Phys.\ Rev.\ Lett.\  {\bf 97} (2006) 021601},
 \arXivOld{hep-ph/0601236}.
\bibitem{Cohen:2006ir}
  A.~G.~Cohen and S.~L.~Glashow,
  ``\textit{A Lorentz-Violating Origin of Neutrino Mass?},''
   \arXivOld{hep-ph/0605036}.
   

 \bibitem{Alfaro:2015fha}
  J.~Alfaro, P.~Gonzalez and R.~Avila,
  ``Electroweak standard model with very special relativity,''
  \doi{10.1103/PhysRevD.91.105007}{Phys.\ Rev.\ D {\bf 91} (2015) 105007},
  \arXiv{1504.04222}{hep-ph}.
  

  
\bibitem{Upadhyay:2016hdj}
S.~Upadhyay and P.~K.~Panigrahi,
``Quantum Gauge Freedom in Very Special Relativity,''
  \doi{10.1016/j.nuclphysb.2016.12.009}{Nucl. Phys. B \textbf{915}, 168-183 (2017)},
  \arXiv{1608.03947}{hep-th}.
 \bibitem{Nayak:2016zed}
  A.~C.~Nayak and P.~Jain,
 ``Phenomenological Implications of Very Special Relativity,''
 \doi{10.1103/PhysRevD.96.075020}{Phys.\ Rev.\ D {\bf 96} (2017) no.7,  075020},
  \arXiv{1610.01826}{hep-ph}.

\bibitem{Alfaro:2017umk}
  J.~Alfaro,
  ``A $Sim(2)$ invariant dimensional regularization,''
  \doi{10.1016/j.physletb.2017.06.018}{Phys.\ Lett.\ B {\bf 772} (2017) 100},
\arXiv{1704.02299}{hep-th}.

        
\bibitem{Bufalo:2017yxs}
R.~Bufalo and S.~Upadhyay,
``Axion Mass Bound in Very Special Relativity,''
 \doi{10.1016/j.physletb.2017.06.070}{Phys. Lett. B \textbf{772}, 420-425 (2017)},
\arXiv{1707.01345}{hep-th}.
 \bibitem{Alfaro:2019koq}
  J.~Alfaro and A.~Soto,
 ``On the photon mass in Very Special Relativity,''
  \doi{10.1103/PhysRevD.100.055029}{Phys.\ Rev.\ D {\bf 100} (2019) no.5,  055029},
\arXiv{1901.08011}{hep-th}.
\bibitem{Upadhyay:2019ghb}
S.~Upadhyay, M.~B.~Shah and P.~A.~Ganai,
``Lorentz-violating gaugeon formalism for rank-2 tensor theory,''
\doi{10.1142/S0217732319502456}{Mod. Phys. Lett. A \textbf{34}, no.30, 1950245 (2019)},
\arXiv{1906.03188}{hep-th}.            

\bibitem{Bufalo:2019kea}
R.~Bufalo and T.~Cardoso e Bufalo,
``Tree-level processes in very special relativity,''
\doi{10.1103/PhysRevD.100.125017}{Phys. Rev. D \textbf{100}, no.12, 125017 (2019)},
\arXiv{1911.08386}{hep-th}.

\bibitem{Banerjee:2001un}
R.~Banerjee and S.~Ghosh,
``Seiberg-Witten map and the axial anomaly in noncommutative field theory,''
\doi{10.1016/S0370-2693(02)01566-6}{Phys. Lett. B \textbf{533}, 162-167 (2002)},
\arXivOld{hep-th/0110177}.
\bibitem{Martin:2005gt}
C.~P.~Martin and C.~Tamarit,
``The U(1)a anomaly in noncommutative SU(N) theories,''
\doi{10.1103/PhysRevD.72.085008}{Phys. Rev. D \textbf{72}, 085008 (2005)},
 \arXivOld{hep-th/0503139}.

\bibitem{Sadooghi:2006sx}
N.~Sadooghi and A.~Jafari Salim,
``Axial anomaly of QED in a strong magnetic field and noncommutative anomaly,''
\doi{10.1103/PhysRevD.74.085032}{Phys. Rev. D \textbf{74}, 085032 (2006)},
 \arXivOld{hep-th/0608112}.


\bibitem{Arias:2007xt}
P.~Arias, H.~Falomir, J.~Gamboa, F.~Mendez and F.~A.~Schaposnik,
``Chiral Anomaly Beyond Lorentz Invariance,''
\doi{10.1103/PhysRevD.76.025019}{Phys. Rev. D \textbf{76}, 025019 (2007)},
\arXiv{0705.3263}{hep-th}.

\bibitem{Salvio:2008ta}
A.~Salvio,
``Relaxing Lorentz invariance in general perturbative anomalies,''
\doi{10.1103/PhysRevD.78.085023}{Phys. Rev. D \textbf{78}, 085023 (2008)},
\arXiv{0809.0184}{hep-th}.
\bibitem{Mariz:2015yua}
T.~Mariz, J.~R.~Nascimento and A.~Y.~Petrov,
``On the Adler-Bell-Jackiw anomaly in a Horava-Lifshitz-like QED,''
\doi{10.1209/0295-5075/112/61002}{EPL \textbf{112}, no.6, 61002 (2015)},
\arXiv{1505.00715}{hep-th}.

\bibitem{AlMasri:2019uzr}
M.~W.~Almasri,
``Axial-anomaly in noncommutative QED and Pauli-Villars regularization,''
\doi{10.1142/S0217751X19501501}{Int. J. Mod. Phys. A \textbf{34}, no.26, 1950150 (2019)},
\arXiv{1909.10280}{hep-th}.


\bibitem{Alfaro:2019snr}
J.~Alfaro and A.~Soto,
``\textit{Schwinger model \`a la Very Special Relativity},''
\doi{10.1016/j.physletb.2019.134923}{Phys. Lett. B \textbf{797}, 134923 (2019)},
\arXiv{1907.06273}{hep-th}.
\bibitem{Alfaro:2020zwo}
J.~Alfaro,
``\textit{Axial anomaly in very special relativity},''
\doi{10.1103/PhysRevD.103.075011}{Phys. Rev. D \textbf{103} (2021) no.7, 075011},
\arXiv{2012.14431}{hep-th}.

\bibitem{Alvarez:2008uy}
E.~Alvarez and R.~Vidal,
``\textit{Very Special (de Sitter) Relativity},''
\doi{10.1103/PhysRevD.77.127702}{Phys. Rev. D \textbf{77} (2008), 127702},
\arXiv{0803.1949}{hep-th}.

\bibitem{Muck:2008bd}
W.~Muck,
``\textit{Very Special Relativity in Curved Space-Times},''
\doi{10.1016/j.physletb.2008.10.028}{Phys. Lett. B \textbf{670} (2008), 95-98},
\arXiv{0806.0737}{hep-th}.
\bibitem{Barcelos-Neto:1985yfn}
J.~Barcelos-Neto and A.~K.~Das,
``\textit{Path Integrals and the Solution of Schwinger Model in Curved Space-time},''
\doi{10.1103/PhysRevD.33.2262}{Phys. Rev. D \textbf{33} (1986), 2262}.

\end{thebibliography}
\end{document}